\documentclass[a4paper]{article}

\usepackage{INTERSPEECH2016}

\usepackage{graphicx}
\usepackage{amssymb,amsmath,bm}
\usepackage{textcomp}
\usepackage{algorithm}
\usepackage{algorithmic}
\usepackage{caption}
\usepackage{booktabs}
\usepackage{multirow}
\usepackage{graphicx,subfigure}

\ninept

\title{
  Audio Word2Vec: Unsupervised Learning of Audio Segment Representations \\
  using Sequence-to-sequence Autoencoder
}

\makeatletter
\def\name#1{\gdef\@name{#1\\}}
\makeatother \name{{\em Yu-An Chung, Chao-Chung Wu, Chia-Hao Shen, Hung-Yi Lee, Lin-Shan Lee}}

\address{College of Electrical Engineering and Computer Science, National Taiwan University \\
  {\small \tt \{b01902040, b01902038, r04921047, hungyilee\}@ntu.edu.tw, lslee@gate.sinica.edu.tw}
}

\begin{document}

\maketitle

\begin{abstract}

  The vector representations of fixed dimensionality for words (in text) offered by Word2Vec have been shown to be very useful in many application scenarios, in particular due to the semantic information they carry.
  This paper proposes a parallel version, the Audio Word2Vec.
  It offers the vector representations of fixed dimensionality for variable-length audio segments.
  These vector representations are shown to describe the sequential phonetic structures of the audio segments to a good degree, with very attractive real world applications such as query-by-example Spoken Term Detection (STD).
  In this STD application, the proposed approach significantly outperformed the conventional Dynamic Time Warping (DTW) based approaches at significantly lower computation requirements.
  We propose unsupervised learning of Audio Word2Vec from audio data without human annotation using Sequence-to-sequence Autoencoder (SA).
  SA consists of two RNNs equipped with Long Short-Term Memory (LSTM) units: the first RNN (encoder) maps the input audio sequence into a vector representation of fixed dimensionality, and the second RNN (decoder) maps the representation back to the input audio sequence.
  The two RNNs are jointly trained by minimizing the reconstruction error.
  Denoising Sequence-to-sequence Autoencoder (DSA) is further proposed offering more robust learning.

\end{abstract}

\section{Introduction}

  Word2Vec~\cite{mikolov2013distributed,mikolov2013efficient,le2014distributed} transforming each word (in text) into a vector of fixed dimensionality has been shown to be very useful in various applications in natural language processing.
  In particular, the vector representations obtained in this way usually carry plenty of semantic information about the word.
  It is therefore interesting to ask a question: can we transform the audio segment of each word into a vector of fixed dimensionality? If yes, what kind of information these vector representations can carry and in what kind of applications can these ``audio version of Word2Vec'' be useful?
  This paper is to try to answer these questions at least partially.

  Representing variable-length audio segments by vectors with fixed dimensionality has been very useful for many speech applications.
  For example, in speaker identification~\cite{IvectorIS09}, audio emotion classification~\cite{EmotionChallengeIS09}, and spoken term detection (STD)~\cite{MyJournal_SVM,segment2vectorIS13,segment2vectorIS12}, the audio segments are usually represented as feature vectors to be applied to standard classifiers to determine the speaker or emotion labels, or whether containing the input queries.
  In query-by-example spoken term detection (STD), by representing each word segment as a vector, easier indexing makes the retrieval much more efficient~\cite{levin2013fixed,SRAILICASSP15,kamper2015deep}.

  But audio segment representation is still an open problem.
  It is common to use i-vectors to represent utterances in speaker identification~\cite{IvectorIS09}, and several approaches have been successfully used in STD~\cite{SRAILICASSP15,MyJournal_SVM,segment2vectorIS13,segment2vectorIS12}.
  But these vector representations may not be able to precisely describe the sequential phonetic structures of the audio segments as we wish to have in this paper, and these approaches were developed primarily in more heuristic ways, rather than learned from data.
  Deep learning has also been used for this purpose~\cite{WordEmbedIS14,QbyELSTMICASSP15}.
  By learning Recurrent Neural Network (RNN) with an audio segment as the input and the corresponding word as the target, the outputs of the hidden layer at the last few time steps can be taken as the representation of the input segment~\cite{QbyELSTMICASSP15}.
  However, this approach is supervised and therefore needs a large amount of labeled training data.

  On the other hand, autoencoder has been a very successful machine learning technique for extracting representations in an unsupervised way~\cite{hinton2006reducing,baldi2012autoencoders}, but its input should be vectors of fixed dimensionality.
  This is a dreadful limitation because audio segments are intrinsically expressed as sequences of arbitrary length.
  A general framework was proposed to encode a sequence using Sequence-to-sequence Autoencoder (SA), in which a RNN is used to encode the input sequence into a fixed-length representation, and then another RNN to decode this input sequence out of that representation.
  This general framework has been applied in natural language processing~\cite{SAforNLP15,SkipThought} and video processing~\cite{srivastava2015unsupervised}, but not yet on speech signals to our knowledge.

  In this paper, we propose to use Sequence-to-sequence Autoencoder (SA) to represent variable-length audio segments by vectors with fixed dimensionality.
  We hope the vector representations obtained in this way can describe more precisely the sequential phonetic structures of the audio signals, so the audio segments that sound alike would have vector representations nearby in the space.
  This is referred to as Audio Word2Vec in this paper.
  Different from the previous work~\cite{WordEmbedIS14,QbyELSTMICASSP15,kamper2015deep}, here learning SA does not need any supervision.
  That is, only the audio segments without human annotation are needed, good for applications in the low resource scenario.
  Inspired from denoising autoencoder~\cite{vincent2008extracting,vincent2010stacked}, an improved version of Denoising Sequence-to-sequence Autoencoder (DSA) is also proposed.
  Among the many possible applications, we chose query-by-example STD in this preliminary study, and show that the proposed Audio Word2Vec can be very useful.
  Query-by-example STD using Audio Word2Vec here is much more efficient than the conventional Dynamic Time Warping (DTW) based approaches, because only the similarities between two single vectors are needed, in additional to the significantly better retrieval performance obtained.

\section{Proposed Approach}

  The goal here is to transform an audio segment represented by a variable-length sequence of acoustic features such as MFCC, $\mathbf{x} = (x_{1}, x_{2}, ..., x_{T})$, where $x_{t}$ is the acoustic feature at time $t$ and $T$ is the length, into a vector representation of fixed dimensionality $\mathbf{z} \in \mathbb{R}^{d}$, where $d$ is the dimensionality of the encoded space.
  It is desired that this vector representation can describe to some degree the sequential phonetic structure of the original audio segment.
  Below we first give a recap on the RNN Encoder-Decoder framework~\cite{sutskever2014sequence,cho2014learning} in Section~\ref{sec:general_seq_to_seq}, followed by the formal presentation of the proposed Sequence-to-sequence Autoencoder (SA) in Section~\ref{sec:sequence_autoencoder}, and its extension in Section~\ref{sec:denoising}.

  \subsection{RNN Encoder-Decoder framework}
    \label{sec:general_seq_to_seq}

    RNNs are neural networks whose hidden neurons form a directed cycle.
    Given a sequence $\mathbf{x} = (x_{1}, x_{2}, ..., x_{T})$, RNN updates its hidden state $\mathbf{h}_{t}$ according to the current input $x_t$ and the previous $\mathbf{h}_{t-1}$.
    The hidden state $\mathbf{h}_{t}$ acts as an internal memory at time $t$ that enables the network to capture dynamic temporal information, and also allows the network to process sequences of variable length.
    In practice, RNN does not seem to learn long-term dependencies~\cite{bengio1994learning}, so LSTM~\cite{hochreiter1997long} has been widely used to conquer such difficulties.
    Because many amazing results were achieved by LSTM, RNN has widely been equipped with LSTM units~\cite{schmidhuber2007training,Wierstra2009b,sak2014long,doetsch2014fast,cho2014learning,chung2014empirical,greff2015lstm}.

    RNN Encoder-Decoder~\cite{cho2014learning,sutskever2014sequence} consists of an Encoder RNN and a Decoder RNN.
    The Encoder RNN reads the input sequence $\mathbf{x} = (x_{1}, x_{2}, ..., x_{T})$ sequentially and the hidden state $\mathbf{h}_{t}$ of the RNN is updated accordingly.
    After the last symbol $x_{T}$ is processed, the hidden state $\mathbf{h}_{T}$ is interpreted as the learned representation of the whole input sequence.
    Then, by taking $\mathbf{h}_{T}$ as input, the Decoder RNN generates the output sequence $\mathbf{y} = (y_{1}, y_{2}, ..., x_{T'})$ sequentially, where $T$ and $T'$ can be different, or the length of $\mathbf{x}$ and $\mathbf{y}$ can be different.
    Such RNN Encoder-Decoder framework is able to handle input of variable length.
    Although there may exist a considerable time lag between the input symbols and their corresponding output symbols, LSTM units in RNN are able to handle such situations well due to the powerfulness of LSTM in modeling long-term dependencies.

  \subsection{Sequence-to-sequence Autoencoder (SA)}
    \label{sec:sequence_autoencoder}

    \begin{figure}[h]
      \vspace{-0.3cm}
      \centering
      \includegraphics[scale=0.27]{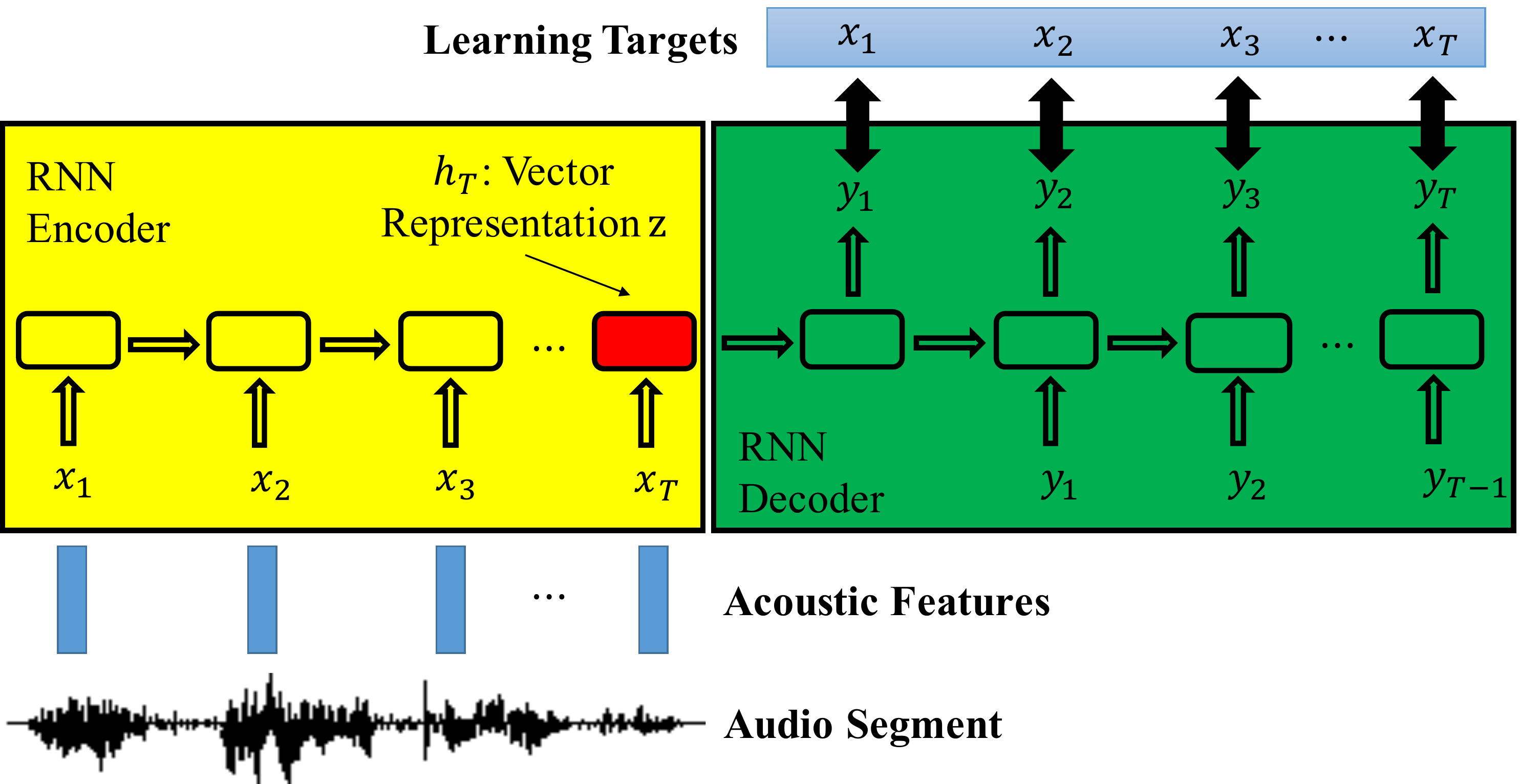}
      \vspace{-0.5cm}
      \caption{
        Sequence-to-sequence Autoencoder (SA) consists of two RNNs: RNN Encoder (the left large block in yellow) and RNN Decoder (the right large block in green).
        The RNN Encoder reads an audio segment represented as an acoustic feature sequence $\mathbf{x} = (x_{1}, x_{2}, ..., x_{T})$ and maps it into a vector representation of fixed dimensionality $\mathbf{z}$; and the RNN Decoder maps the vector $\mathbf{z}$ to another sequence $\mathbf{y} = (y_{1}, y_{2}, ..., y_{T})$.
        The RNN Encoder and Decoder are jointly trained to make the output sequence $\mathbf{y}$ as close to the input sequence $\mathbf{x}$ as possible, or to minimize the reconstruction error.
      }
      \vspace{-0.45cm}
      \label{fig:sequence_autoencoder}
    \end{figure}

    Figure~\ref{fig:sequence_autoencoder} depicts the structure of the proposed Sequence-to-sequence Autoencoder (SA), which integrates the RNN Encoder-Decoder framework with Autoencoder for unsupervised learning of audio segment representations.
    SA consists of an Encoder RNN (the left part of Figure~\ref{fig:sequence_autoencoder}) and a Decoder RNN (the right part).
    Given an audio segment represented as an acoustic feature sequence $\mathbf{x} = (x_{1}, x_{2}, ..., x_{T})$ of any length $T$, the Encoder RNN reads each acoustic feature $x_{t}$ sequentially and the hidden state $\mathbf{h}_{t}$ is updated accordingly.
    After the last acoustic feature $x_{T}$ has been read and processed, the hidden state $\mathbf{h}_{T}$ of the Encoder RNN is viewed as the \emph{learned representation} $\mathbf{z}$ of the input sequence (the small red block in Figure~\ref{fig:sequence_autoencoder}).
    The Decoder RNN takes $\mathbf{h}_{T}$ as the first input and generates its first output $y_{1}$.
    Then it takes $y_{1}$ as input to generate $y_{2}$, and so on.
    Based on the principles of Autoencoder~\cite{hinton2006reducing,baldi2012autoencoders}, the target of the output sequence $\mathbf{y} = (y_{1}, y_{2}, ..., y_{T})$ is the input sequence $\mathbf{x} = (x_{1}, x_{2}, ..., x_{T})$.
    In other words, the RNN Encoder and Decoder are jointly trained by minimizing the reconstruction error, measured by the general mean squared error $\sum_{t=1}^{T}\| x_{t} - y_{t} \|^2$.
    Because the input sequence is taken as the learning target, the training process does not need any labeled data.
    The fixed-length vector representation $\mathbf{z}$ will be a meaningful representation for the input audio segment $\mathbf{x}$, because the whole input sequence $\mathbf{x}$ can be reconstructed from $\mathbf{z}$ by the RNN Decoder.

  \subsection{Denoising Sequence-to-sequence Autoencoder (DSA)}
    \label{sec:denoising}

    To learn more robust representation, we further apply the denoising criterion~\cite{vincent2010stacked} to the SA learning proposed above.
    The input acoustic feature sequence $\mathbf{x}$ is randomly added with some noise, yielding a \emph{corrupted} version $\tilde{\mathbf{x}}$.
    Here the input to SA is $\tilde{\mathbf{x}}$, and SA is expected to generate the output $\mathbf{y}$ closest to the original $\mathbf{x}$ based on $\tilde{\mathbf{x}}$.
    The SA learned with this denoising criterion is referred to as Denoising SA (DSA) below.
    \vspace{-0.2cm}

\section{An Example Application: Query-by-example STD}
  \label{sec:qbe}
  \begin{figure}[h]
    \vspace{-0.5cm}
    \includegraphics[scale=0.25]{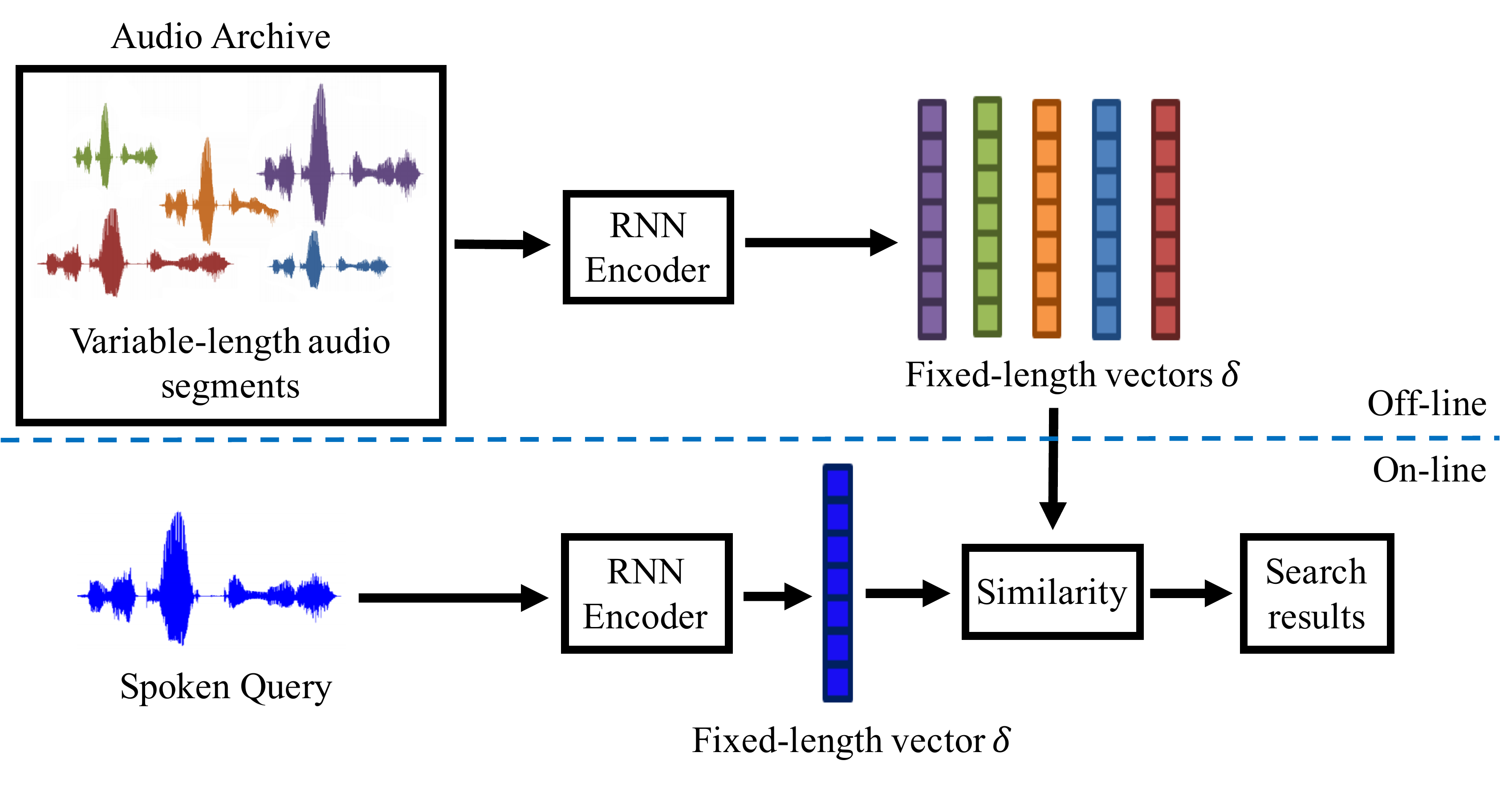}
    \vspace{-0.5cm}
    \caption{
      The example application of query-by-example STD.
      All audio segments in the audio archive are segmented based on word boundaries and represented by fixed-length vectors off-line.
      When a spoken query is entered, it is also represented as a vector, and the similarity between this vector and all vectors for segments in the archive are calculated, and the audio segments are ranked accordingly.
    }
    \vspace{-0.35cm}
    \label{fig:qbe}
  \end{figure}

  The audio segment representation $\mathbf{z}$ learned in the last section can be applied in many possible scenarios.
  Here in the preliminary tests we consider the unsupervised query-by-example STD, whose target is to locate the occurrence regions of the input spoken query term in a large spoken archive without speech recognition.
  Figure~\ref{fig:qbe} shows how the representation $\mathbf{z}$ proposed here can be easily used in this task.
  This approach is inspired from the previous work~\cite{SRAILICASSP15}, but completely different in the ways to represent the audio segments.
  In the upper half of Figure~\ref{fig:qbe}, the audio archive are segmented based on word boundaries into variable-length sequences, and then the system exploits the trained RNN encoder in Figure~\ref{fig:sequence_autoencoder} to encode these audio segments into fixed-length vectors.
  All these are done off-line.
  In the lower left corner of Figure~\ref{fig:qbe}, when a spoken query is entered, the input spoken query is similarly encoded by the same RNN encoder into a vector.
  The system then returns a list of audio segments in the archive ranked according to the cosine similarities evaluated between the vector representation of the query and those of all segments in the archive.
  Note that the computation requirements for the on-line process here are extremely low.
  \vspace{-0.3cm}

\section{Experiments}

  Here, we report the experiments and results including the experimental setup (Section~\ref{sec:exp_setup}), analysis for the vector representations learned (Section~\ref{sec:analysis}), and query-by-example STD results (Section~\ref{sec:qbe_std}).
  \vspace{-0.2cm}

  \subsection{Experimental Setup}
    \label{sec:exp_setup}

    We used LibriSpeech corpus~\cite{vassil2015librispeech} as the data for the experiments.
    Due to the computation resource limitation, only 5.4 hours \textit{dev-clean} (produced by 40 speakers) was used for training SA and DSA.
    Another 5.4 hours \textit{test-clean} (produced by a completely different group of 40 speakers) was the testing set.
    MFCCs of 13-dim were used as the acoustic features.
    Both the training and testing sets were segmented according to the word boundaries obtained by forced alignment with respect to the reference transcriptions.
    Although the oracle word boundaries were used here for the query-by-example STD in the preliminary tests, the comparison in Section~\ref{sec:qbe_std} was fair since the baselines used the same segmentation.

    SA and DSA were implemented with Theano~\cite{bergstra+al:2010-scipy,DBLP:journals/corr/abs-1211-5590}.
    The network structure and hyper parameters were set as below without further tuning if not specified: 
    \begin{itemize}
      \item{
        Both RNN Encoder and Decoder consisted of one hidden layer with 100 LSTM units.
        That is, each input segment was mapped into a 100-dim vector representation by SA or DSA.
        We adopted the LSTM described in~\cite{DBLP:journals/corr/Graves13}, and the \emph{peephole connections}~\cite{gers2000recurrent} were added.
      }
      \item{
        The networks were trained by SGD without momentum, with a fixed learning rate of 0.3 and 500 epochs.
      }
      \item{
        For DSA, zero-masking~\cite{vincent2010stacked} was used to generate the noisy input $\tilde{\mathbf{x}}$, which randomly wiped out some elements in each input acoustic feature and set them to zero.
        The probability to be wiped out was set to 0.3.
      }
    \end{itemize}

  For query-by-example STD, the testing set served as the audio archive to be searched through.
  Each word segment in the testing set was taken as a spoken query once.
  When a word segment was taken as a query, it was excluded from the archive during retrieval.
  There were 5557 queries in total.
  Mean Average Precision (MAP)~\cite{MAP_PRcurve} was used as the evaluation measure for query-by-example STD as the previous work~\cite{Chung2014icassp}.

  \subsection{Analysis of the Learned Representations}
    \label{sec:analysis}

    In this section, we wish to verify that the vector representations obtained here describe the sequential phonetic structures of the audio segments, or words with similar pronunciations have close vector representations.

    We first used the SA and DSA learned from the training set to encode the segments in the testing set, which were never seen during training, and then computed the cosine similarity between each segment pair.
    The average results for groups of pairs clustered by the phoneme edit distances between the two words in the pair are in Table~\ref{tab:edit_distance}.
    It is clear from Table~\ref{tab:edit_distance} that two segments for words with larger phoneme sequence edit distances have obviously smaller cosine similarity in average.
    We also found that SA and DSA can even very clearly distinguish those word segments with only one different phoneme, because the average similarity for the cluster of zero edit distance is 0 (exactly the same pronunciation) is remarkably larger than that for edit distance being 1.
    The gap is even larger for clusters with edit distances being 2 vs. 1.
    For the cluster of zero edit distance, the average similarity is only 0.48-0.50.
    This is reasonable because it is well-known that even with exactly identical phoneme segments, the acoustic realizations can be very different.
    It seems DSA is slightly better than SA, because the similarity by DSA for pairs with 0 or 1 edit distances is slightly larger, while much smaller for those with 4, 5 or more edit distances.
    These results show that SA and DSA can in fact encode the acoustic signals into vector representations describing the sequential phonetic structures of the signals.
    The performance between SA and DSA can be further compared with other baselines in Section~\ref{sec:qbe_std}.

    \begin{table}[h!]
      \centering
      \caption{
        The average cosine similarity between the vector presentations for all the segment pairs in the testing set, clustered by the phoneme sequence edit distances.
      }
      \vspace{-0.3cm}
      \label{tab:edit_distance}
      \begin{tabular}{|c|c|c|c|}
      \hline
      \multirow{2}{*}{
        \begin{tabular}[c]{@{}c@{}}Phoneme Sequence Edit Distance\end{tabular}} & \multirow{2}{*}{\# of pairs} & \multicolumn{2}{c|}{\begin{tabular}[c]{@{}c@{}}Average cosine\\ similarity\end{tabular}} \\ \cline{3-4} 
                        &                  &           SA         &          DSA        \\ \hline
        0 (same)        &       95222      &        0.4847        &        0.5012       \\ \hline	
        1               &       95355      &        0.4016        &        0.4237       \\ \hline	
        2               &      987934      &        0.2674        &        0.2798       \\ \hline	
        3               &     4651318      &        0.0835        &        0.0986       \\ \hline	
        4               &     4791482      &        0.0255        &        0.0196       \\ \hline	
        5 or more       &     3078684      &        0.0051        &        0.0013       \\ \hline	
      \end{tabular}
    \end{table}

    \vspace{-0.3cm}
    Because RNN Encoder in Figure~\ref{fig:sequence_autoencoder} reads the input acoustic sequence sequentially, it may be possible that the last few acoustic features dominate the vector representation, and as a result those words with the same suffixes are hard to be distinguished by the learned representations.
    To justify this hypothesis, we analyze the following two sets of words with the same suffixes:
    \begin{itemize}
      \item Set 1: father, mother, another
      \item Set 2: ever, never, however
    \end{itemize}
    Same as Table~\ref{tab:edit_distance}, we used cosine similarity between the two encoded vectors averaged over each group of pairs to measure the differences, with results in Table~\ref{tab:series_cosine}.
    \begin{table}[h!]
      \centering
      \caption{
        The average cosine similarity between the vector representations for words with the same suffixes.
      }
      \label{tab:series_cosine}
      \begin{tabular}{|c|c|c|c|}
      \hline
      \multirow{2}{*}{\textbf{Set 1}} & \multirow{2}{*}{\# of pairs} & \multicolumn{2}{c|}{Average cosine similarity} \\ \cline{3-4} 
                               		  &                              &    \hspace{0.3cm}    SA    \hspace{0.3cm}          &    DSA          \\ \hline
        father vs. father     		  &              406             &    \hspace{0.3cm}  0.5047  \hspace{0.3cm}          &   0.5209        \\ \hline
        father vs. mother     		  &              832             &    \hspace{0.3cm}  0.2639  \hspace{0.3cm}          &   0.2748        \\ \hline
        father vs. another    		  &             1276             &    \hspace{0.3cm}  0.1964  \hspace{0.3cm}          &   0.1861        \\ \hline
      \multirow{2}{*}{\textbf{Set 2}} & \multirow{2}{*}{\# of pairs} & \multicolumn{2}{c|}{Average cosine similarity} \\ \cline{3-4} 
                       				  &                              &    \hspace{0.3cm}    SA    \hspace{0.3cm}          &    DSA          \\ \hline
        ever vs. ever   	          &              46              &    \hspace{0.3cm}  0.4027  \hspace{0.3cm}          &   0.4265        \\ \hline
        ever vs. never                &             1581             &    \hspace{0.3cm}  0.2475  \hspace{0.3cm}          &   0.2238        \\ \hline
        ever vs. however      	      &              713             &    \hspace{0.3cm}  0.2638  \hspace{0.3cm}          &   0.2419        \\ \hline
      \end{tabular}
    \end{table}
    For Set~1 in Table~\ref{tab:series_cosine}, we see the vector representations of the audio segments of ``father'' are much closer to each other than to those for ``mother'', while those for ``another'' are even much farther away.
    Similar for Set~2.
    There exist much more similar examples in the dataset, but we just show two here due to the space limitation.
    These results show that although SA or DSA read the acoustic signals sequentially, words with the same suffix are actually clearly distinguishable from the learned vectors.

\begin{figure}[ht!]
      \begin{center}
        \vspace{-0.4cm}
        \subfigure[Set 3: the first phoneme changes from f to n.]{
            \label{fig:phoneme_diff_a}
            \hspace{-0.7cm} \includegraphics[width=0.5\textwidth]{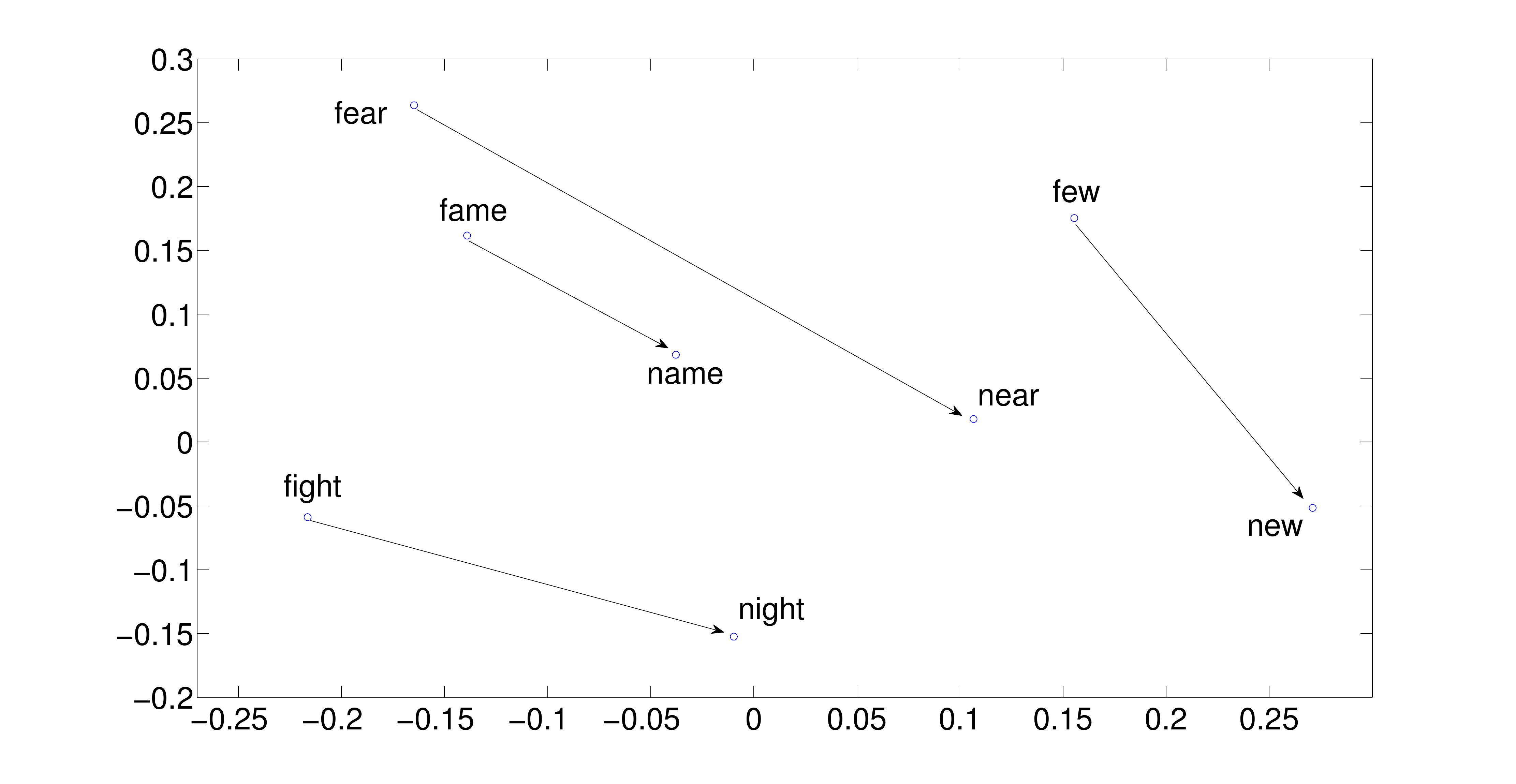}
        } \\
        \vspace{-0.4cm}
        \subfigure[Set 4: the last phoneme differs by existing s or not.]{
           \label{fig:phoneme_diff_b}
            \hspace{-0.7cm} \includegraphics[width=0.5\textwidth]{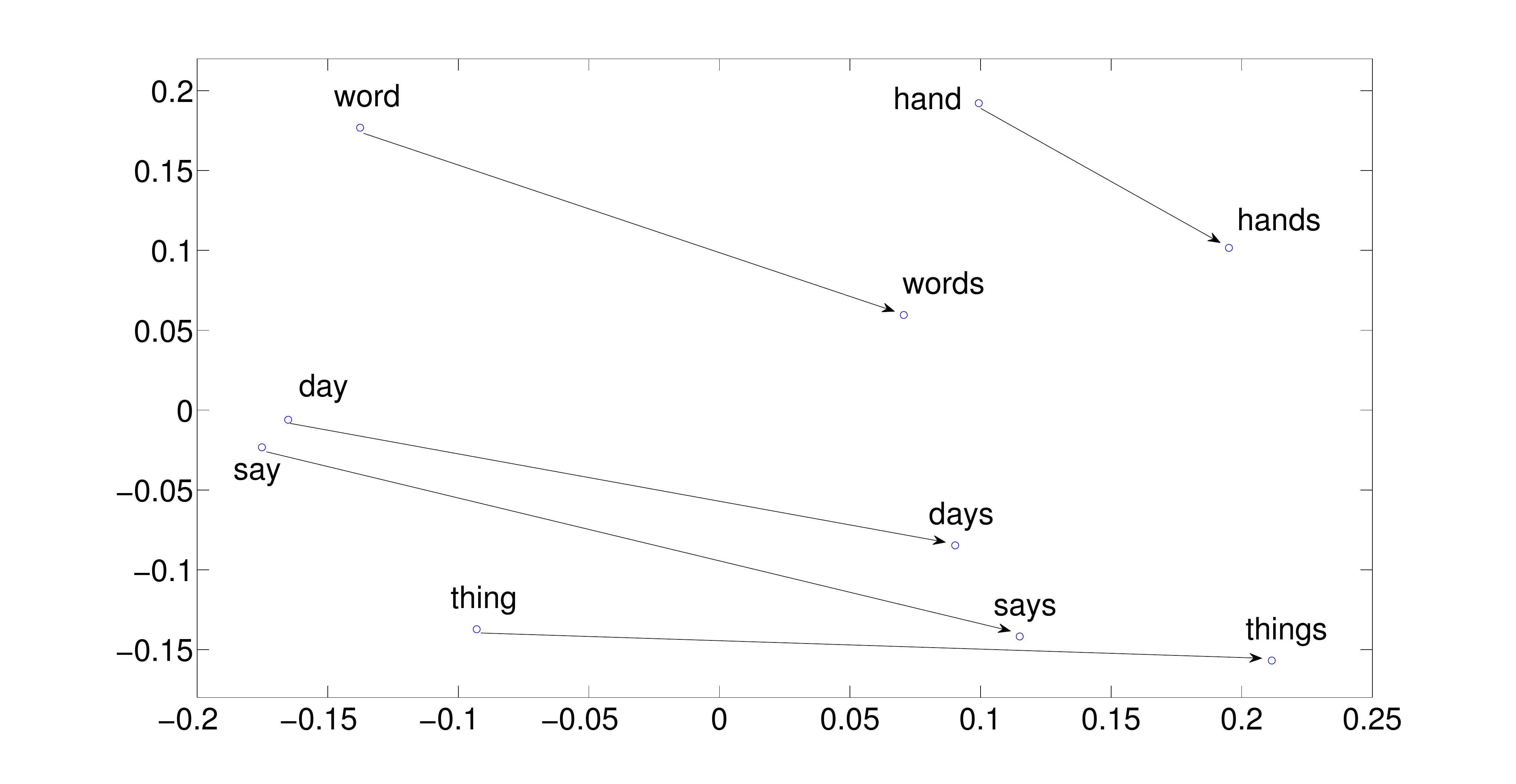}
        }
      \end{center}
      \vspace{-0.6cm}
      \caption{
        Difference vectors between the average vector representations for word pairs differing by (a) the first phoneme as in Set 3 and (b) the last phoneme as in Set 4.
      }
      \label{fig:phoneme_diff}
    \end{figure}

    For another test, we selected two sets of word pairs differing by only the first phoneme or the last phoneme as below.
    \begin{itemize}
      \item Set 3: (new, few), (near, fear), (name, fame), (night, fight)
      \item Set 4: (hand, hands), (word, words), (day, days), (say, says), (thing, things)
    \end{itemize}
    We evaluated the difference vectors between the averages of the vector representations by SA for the two words in a pair, e.g., $\delta(\mathrm{word1}) - \delta(\mathrm{word2})$, and reduced the dimensionality of the difference vectors to 2 to see more easily how they actually differentiated in the vector space~\cite{mikolov2013distributed}.
    The results for Sets 3 and 4 are respectively in Figure~\ref{fig:phoneme_diff_a} and Figure~\ref{fig:phoneme_diff_b}.
    We see in each case the difference vectors are in fact close in their directions and magnitudes, for example, $\delta(\mathrm{new}) - \delta(\mathrm{few})$ is close to $\delta(\mathrm{night}) - \delta(\mathrm{fight})$ in Figure~\ref{fig:phoneme_diff_a}, and $\delta(\mathrm{days}) - \delta(\mathrm{day})$ is close to $\delta(\mathrm{things}) - \delta(\mathrm{thing})$ in Figure~\ref{fig:phoneme_diff_b}.
    This implies ``phoneme replacement'' is somehow realizable here, for example, $\delta (\mathrm{new}) - \delta(\mathrm{few}) + \delta(\mathrm{fight})$ is not very far from $\delta(\mathrm{fight})$ in Figure~\ref{fig:phoneme_diff_a}, and $\delta(\mathrm{things}) - \delta(\mathrm{thing}) + \delta(\mathrm{day})$ is not very far from $\delta(\mathrm{days})$ in Figure~\ref{fig:phoneme_diff_b}.
    They are not ``very close'' because it is well-known that the same words can have very different acoustic realizations.

    \vspace{-0.5cm}
    All results above show the vector representations in test offer very good descriptions for the sequential phonemic structures of the acoustic segments.
    \vspace{-0.2cm}

  \subsection{Query-by-example STD}
    \label{sec:qbe_std}

    \vspace{-0.1cm}
    Here we compared the performance of using the vector representations from SA and DSA in query-by-example STD with two baselines.
    The first baseline used frame-based DTW~\cite{sakoe1978dynamic} which is commonly used in query-by-example STD tasks.
    Here we used the vanilla version of DTW (with Euclidean distance as the frame-level distance) to compute the similarities between the input query and the audio segments, and ranked the segments accordingly.
    The second baseline used the same framework in Figure~\ref{fig:qbe}, except that the RNN Encoder is replaced by a manually designed encoder.
    In this encoder, we first divide the input acoustic feature sequence $\mathbf{x} = (x_{1}, x_{2}, ..., x_{T})$, where $x_{t}$ is the 13-dimensional acoustic feature vector at time $t$, into $m$ segments with roughly equal length $\frac{T}{m}$, then average each segment of vectors into a single 13-dimensional vector, and finally concatenate these $m$ average vectors sequentially into a vector representation of dimensionality $13 \times m$.
    We refer to this approach as Na{\"i}ve Encoder (NE) below.
    Although NE is simple, similar approaches have been used in STD and achieved successful results~\cite{MyJournal_SVM,segment2vectorIS13,segment2vectorIS12}.
    We found empirically that NE with 52, 78 and 104 dimensions (or $m = 4, 6, 8$), represented by $\mathrm{NE}_{52}$, $\mathrm{NE}_{78}$, and $\mathrm{NE}_{104}$ below, yielded the best results. So $\mathrm{NE}_{52}$, $\mathrm{NE}_{78}$, and $\mathrm{NE}_{104}$ were used in the following experiments.
    \vspace{-0.4cm}

    \begin{figure}[h]
      \centering
      \hspace{-0.6cm} \includegraphics[scale=0.23]{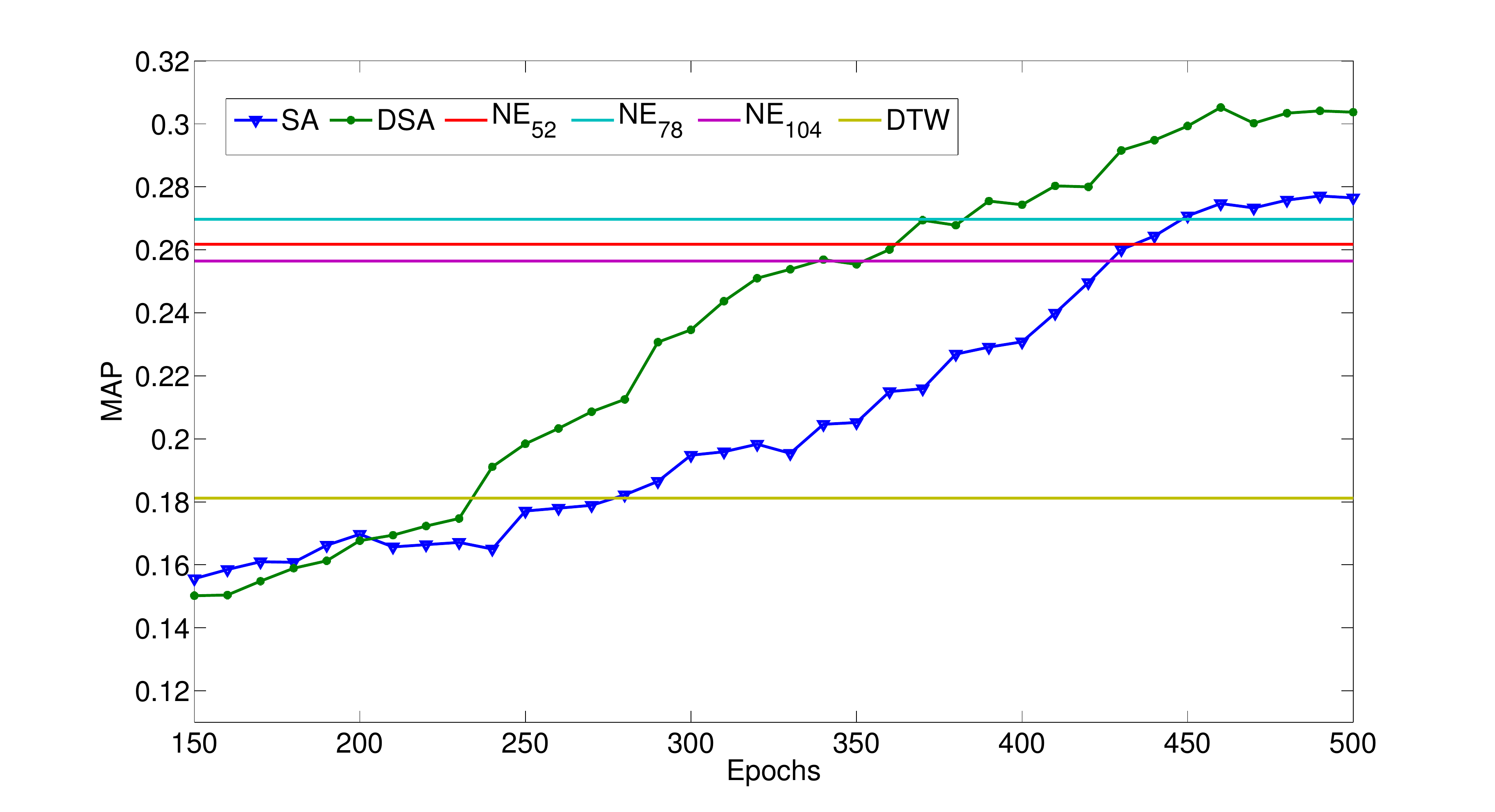}
      \vspace{-0.6cm}
      \caption[font=small]{
        The retrieval performance in MAP of DTW, Na{\"i}ve Encoder ($\mathrm{NE}_{52}$ , $\mathrm{NE}_{78}$, and $\mathrm{NE}_{104}$), SA and DSA.
        The horizontal axis is the number of epochs in training SA and DSA, not relevant to all baselines.
      }
      \label{fig:learning_curve}
      \vspace{-0.3cm}
    \end{figure}
    Figure~\ref{fig:learning_curve} shows the retrieval performance in MAP of DTW, $\mathrm{NE}_{52}$, $\mathrm{NE}_{78}$, $\mathrm{NE}_{104}$, SA and DSA.
    The horizontal axis is the number of epochs in training SA and DSA, not relevant to all baselines.
    Besides DTW which computes the distance between audio segments directly, all other approaches map the variable-length audio segments to fixed-length vectors for similarity evaluation, but in different approaches.
    DTW is not comparable with $\mathrm{NE}_{52}$, $\mathrm{NE}_{78}$, and $\mathrm{NE}_{104}$ probably because only the vanilla version was used, and the averaging process in NE may smooth some of the local signal variations which may cause disturbances in DTW.
    We see as the training epochs increased, both SA and DSA resulted in higher MAP than NE.
    SA outperformed $\mathrm{NE}_{52}$, $\mathrm{NE}_{78}$, and $\mathrm{NE}_{104}$ after about 450 epochs, while DSA did that after about 390 epochs, and was able to achieve much higher MAP.
    Here DSA clearly outperformed SA in most cases.
    These results verified that the learned vector representations can be useful in real world applications.

\section{Conclusions and Future Work}

  We propose to use Sequence-to-sequence Autoencoder (SA) and its extension for unsupervised learning of Audio Word2Vec to obtain vector representations for audio segments.
  We show SA can learn vector representations describing the sequential phonetic structures of the audio segments, and these representations can be useful in real world applications, such as query-by-example STD tested in this preliminary study.
  For the future work, we are training the SA on larger corpora with different dimensionality in different extensions, and considering other application scenarios.

\newpage
\eightpt

\bibliographystyle{IEEEtran}
\bibliography{interspeech-16}

\begin{thebibliography}{10}
\providecommand{\url}[1]{#1}
\csname url@samestyle\endcsname
\providecommand{\newblock}{\relax}
\providecommand{\bibinfo}[2]{#2}
\providecommand{\BIBentrySTDinterwordspacing}{\spaceskip=0pt\relax}
\providecommand{\BIBentryALTinterwordstretchfactor}{4}
\providecommand{\BIBentryALTinterwordspacing}{\spaceskip=\fontdimen2\font plus
\BIBentryALTinterwordstretchfactor\fontdimen3\font minus
  \fontdimen4\font\relax}
\providecommand{\BIBforeignlanguage}[2]{{%
\expandafter\ifx\csname l@#1\endcsname\relax
\typeout{** WARNING: IEEEtran.bst: No hyphenation pattern has been}%
\typeout{** loaded for the language `#1'. Using the pattern for}%
\typeout{** the default language instead.}%
\else
\language=\csname l@#1\endcsname
\fi
#2}}
\providecommand{\BIBdecl}{\relax}
\BIBdecl

\bibitem{mikolov2013distributed}
T.~Mikolov, I.~Sutskever, K.~Chen, G.~S. Corrado, and J.~Dean, ``Distributed
  representations of words and phrases and their compositionality,'' in
  \emph{NIPS}, 2013.

\bibitem{mikolov2013efficient}
T.~Mikolov, K.~Chen, G.~Corrado, and J.~Dean, ``Efficient estimation of word
  representations in vector space,'' \emph{arXiv preprint arXiv:1301.3781},
  2013.

\bibitem{le2014distributed}
Q.~V. Le and T.~Mikolov, ``Distributed representations of sentences and
  documents,'' \emph{arXiv preprint arXiv:1405.4053}, 2014.

\bibitem{IvectorIS09}
N.~Dehak, R.~Dehak, P.~Kenny, N.~Brummer, P.~Ouellet, and P.~Dumouchel,
  ``Support vector machines versus fast scoring in the low-dimensional total
  variability space for speaker verification,'' in \emph{INTERSPEECH}, 2009.

\bibitem{EmotionChallengeIS09}
B.~Schuller, S.~Steidl, and A.~Batliner, ``The {INTERSPEECH} 2009 emotion
  challenge,'' in \emph{INTERSPEECH}, 2009.

\bibitem{MyJournal_SVM}
H.-Y. Lee and L.-S. Lee, ``Enhanced spoken term detection using support vector
  machines and weighted pseudo examples,'' \emph{Audio, Speech, and Language
  Processing, IEEE Transactions on}, vol.~21, no.~6, pp. 1272--1284, 2013.

\bibitem{segment2vectorIS13}
I.-F. Chen and C.-H. Lee, ``A hybrid {HMM}/{DNN} approach to keyword spotting
  of short words,'' in \emph{INTERSPEECH}, 2013.

\bibitem{segment2vectorIS12}
A.~Norouzian, A.~Jansen, R.~Rose, and S.~Thomas, ``Exploiting discriminative
  point process models for spoken term detection,'' in \emph{INTERSPEECH},
  2012.

\bibitem{levin2013fixed}
K.~Levin, K.~Henry, A.~Jansen, and K.~Livescu, ``Fixed-dimensional acoustic
  embeddings of variable-length segments in low-resource settings,'' in
  \emph{ASRU}, 2013.

\bibitem{SRAILICASSP15}
K.~Levin, A.~Jansen, and B.~Van~Durme, ``Segmental acoustic indexing for zero
  resource keyword search,'' in \emph{ICASSP}, 2015.

\bibitem{kamper2015deep}
H.~Kamper, W.~Wang, and K.~Livescu, ``Deep convolutional acoustic word
  embeddings using word-pair side information,'' in \emph{ICASSP}, 2016.

\bibitem{WordEmbedIS14}
S.~Bengio and G.~Heigold, ``Word embeddings for speech recognition,'' in
  \emph{INTERSPEECH}, 2014.

\bibitem{QbyELSTMICASSP15}
G.~Chen, C.~Parada, and T.~N. Sainath, ``Query-by-example keyword spotting
  using long short-term memory networks,'' in \emph{ICASSP}, 2015.

\bibitem{hinton2006reducing}
G.~E. Hinton and R.~R. Salakhutdinov, ``Reducing the dimensionality of data
  with neural networks,'' \emph{Science}, vol. 313, no. 5786, pp. 504--507,
  2006.

\bibitem{baldi2012autoencoders}
P.~Baldi, ``Autoencoders, unsupervised learning, and deep architectures,''
  \emph{Unsupervised and Transfer Learning Challenges in Machine Learning,
  Volume 7}, p.~43, 2012.

\bibitem{SAforNLP15}
J.~Li, M.-T. Luong, and D.~Jurafsky, ``A hierarchical neural autoencoder for
  paragraphs and documents,'' in \emph{arXiv preprint arXiv: 1506.01057}, 2015.

\bibitem{SkipThought}
R.~Kiros, Y.~Zhu, R.~Salakhutdinov, R.~S. Zemel, A.~Torralba, R.~Urtasun, and
  S.~Fidler, ``Skip-thought vectors,'' in \emph{arXiv preprint arXiv:
  1506.06726}, 2015.

\bibitem{srivastava2015unsupervised}
N.~Srivastava, E.~Mansimov, and R.~Salakhutdinov, ``Unsupervised learning of
  video representations using {LSTM}s,'' \emph{arXiv preprint
  arXiv:1502.04681}, 2015.

\bibitem{vincent2008extracting}
P.~Vincent, H.~Larochelle, Y.~Bengio, and P.-A. Manzagol, ``Extracting and
  composing robust features with denoising autoencoders,'' in \emph{ICML},
  2008.

\bibitem{vincent2010stacked}
P.~Vincent, H.~Larochelle, I.~Lajoie, Y.~Bengio, and P.-A. Manzagol, ``Stacked
  denoising autoencoders: Learning useful representations in a deep network
  with a local denoising criterion,'' \emph{JMLR}, vol.~11, pp. 3371--3408,
  2010.

\bibitem{sutskever2014sequence}
I.~Sutskever, O.~Vinyals, and Q.~V. Le, ``Sequence to sequence learning with
  neural networks,'' in \emph{NIPS}, 2014.

\bibitem{cho2014learning}
K.~Cho, B.~Van~Merri{\"e}nboer, C.~Gulcehre, D.~Bahdanau, F.~Bougares,
  H.~Schwenk, and Y.~Bengio, ``Learning phrase representations using rnn
  encoder-decoder for statistical machine translation,'' \emph{arXiv preprint
  arXiv:1406.1078}, 2014.

\bibitem{bengio1994learning}
Y.~Bengio, P.~Simard, and P.~Frasconi, ``Learning long-term dependencies with
  gradient descent is difficult,'' \emph{Neural Networks, IEEE Transactions
  on}, vol.~5, no.~2, pp. 157--166, 1994.

\bibitem{hochreiter1997long}
S.~Hochreiter and J.~Schmidhuber, ``Long short-term memory,'' \emph{Neural
  Computation}, vol.~9, no.~8, pp. 1735--1780, 1997.

\bibitem{schmidhuber2007training}
J.~Schmidhuber, D.~Wierstra, M.~Gagliolo, and F.~Gomez, ``Training recurrent
  networks by evolino,'' \emph{Neural Computation}, vol.~19, no.~3, pp.
  757--779, 2007.

\bibitem{Wierstra2009b}
J.~Bayer, D.~Wierstra, J.~Togelius, and J.~Schmidhuber, ``Evolving memory cell
  structures for sequence learning,'' in \emph{ICANN}, 2009.

\bibitem{sak2014long}
H.~Sak, A.~Senior, and F.~Beaufays, ``Long short-term memory recurrent neural
  network architectures for large scale acoustic modeling,'' in
  \emph{INTERSPEECH}, 2014.

\bibitem{doetsch2014fast}
P.~Doetsch, M.~Kozielski, and H.~Ney, ``Fast and robust training of recurrent
  neural networks for offline handwriting recognition,'' in \emph{ICFHR}, 2014.

\bibitem{chung2014empirical}
J.~Chung, C.~Gulcehre, K.~Cho, and Y.~Bengio, ``Empirical evaluation of gated
  recurrent neural networks on sequence modeling,'' \emph{arXiv preprint
  arXiv:1412.3555}, 2014.

\bibitem{greff2015lstm}
K.~Greff, R.~K. Srivastava, J.~Koutn{\'\i}k, B.~R. Steunebrink, and
  J.~Schmidhuber, ``Lstm: A search space odyssey,'' \emph{arXiv preprint
  arXiv:1503.04069}, 2015.

\bibitem{vassil2015librispeech}
V.~Panayotov, G.~Chen, D.~Povey, and S.~Khudanpur, ``Librispeech: an {ASR}
  corpus based on public domain audio books,'' in \emph{ICASSP}, 2015.

\bibitem{bergstra+al:2010-scipy}
J.~Bergstra, O.~Breuleux, F.~Bastien, P.~Lamblin, R.~Pascanu, G.~Desjardins,
  J.~Turian, D.~Warde-Farley, and Y.~Bengio, ``Theano: a {CPU} and {GPU} math
  expression compiler,'' in \emph{SciPy}, 2010.

\bibitem{DBLP:journals/corr/abs-1211-5590}
F.~Bastien, P.~Lamblin, R.~Pascanu, J.~Bergstra, I.~J. Goodfellow, A.~Bergeron,
  N.~Bouchard, D.~Warde{-}Farley, and Y.~Bengio, ``Theano: new features and
  speed improvements,'' \emph{CoRR}, vol. abs/1211.5590, 2012.

\bibitem{DBLP:journals/corr/Graves13}
A.~Graves, ``Generating sequences with recurrent neural networks,''
  \emph{CoRR}, vol. abs/1308.0850, 2013.

\bibitem{gers2000recurrent}
F.~Gers, J.~Schmidhuber \emph{et~al.}, ``Recurrent nets that time and count,''
  in \emph{IJCNN}, 2000.

\bibitem{MAP_PRcurve}
C.~D. Manning, P.~Raghavan, and H.~Sch{\"u}tze, \emph{Introduction to
  information retrieval}.\hskip 1em plus 0.5em minus 0.4em\relax Cambridge
  University Press, 2008.

\bibitem{Chung2014icassp}
C.-T. Chung, C.-A. Chan, and L.-S. Lee, ``Unsupervised spoken term detection
  with spoken queries by multi-level acoustic patterns with varying model
  granularity,'' in \emph{ICASSP}, 2014.

\bibitem{sakoe1978dynamic}
H.~Sakoe and S.~Chiba, ``Dynamic programming algorithm optimization for spoken
  word recognition,'' \emph{Acoustics, Speech and Signal Processing, IEEE
  Transactions on}, vol.~26, no.~1, pp. 43--49, 1978.

\end{thebibliography}

\end{document}